# A Graph Algorithmic Approach to Separate Direct from Indirect Neural Interactions

**Patricia Wollstadt[1] \*, Ulrich Meyer[2], Michael Wibral[1]**

**1** MEG Unit, Brain Imaging Center, Goethe University, Frankfurt/Main, Germany, **2** Institute for Computer Science, Goethe University, Frankfurt/Main, Germany

\* patricia.wollstadt@stud.uni-frankfurt.de

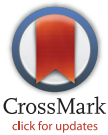

**Data Availability Statement:** The reference MATLAB implementation of the algorithm described in the article is available from the GitHub repository at https://github.com/trentool/TRENTOOL3. Magnetoencephalography data are available from the authors for researchers who meet the criteria for access to confidential data imposed by the ethics committee at Johann Wolfgang Goethe University, Frankfurt, Germany.

**Funding:** MW and PW received financial support from the grant LOEWE "Neuronale Koordination Forschungsschwerpunkt Frankfurt (NeFF)" awarded by the Landes-Offensive zur Entwicklung

## Abstract

Network graphs have become a popular tool to represent complex systems composed of many interacting subunits; especially in neuroscience, network graphs are increasingly used to represent and analyze functional interactions between multiple neural sources. Interactions are often reconstructed using pairwise bivariate analyses, overlooking the multivariate nature of interactions: it is neglected that investigating the effect of one source on a target necessitates to take all other sources as potential nuisance variables into account; also combinations of sources may act jointly on a given target. Bivariate analyses produce networks that may contain spurious interactions, which reduce the interpretability of the network and its graph metrics. A truly multivariate reconstruction, however, is computationally intractable because of the combinatorial explosion in the number of potential interactions. Thus, we have to resort to approximative methods to handle the intractability of multivariate interaction reconstruction, and thereby enable the use of networks in neuroscience. Here, we suggest such an approximative approach in the form of an algorithm that extends fast bivariate interaction reconstruction by identifying potentially spurious interactions post-hoc: the algorithm uses interaction delays reconstructed for directed bivariate interactions to tag potentially spurious edges on the basis of their timing signatures in the context of the surrounding network. Such tagged interactions may then be pruned, which produces a statistically conservative network approximation that is guaranteed to contain non-spurious interactions only. We describe the algorithm and present a reference implementation in MATLAB to test the algorithm's performance on simulated networks as well as networks derived from magnetoencephalographic data. We discuss the algorithm in relation to other approximative multivariate methods and highlight suitable application scenarios. Our approach is a tractable and data-efficient way of reconstructing approximative networks of multivariate interactions. It is preferable if available data are limited or if fully multivariate approaches are computationally infeasible.

## Introduction

Complex systems are often composed of many interacting simpler subunits. To summarize our knowledge about such a system in an accessible format we frequently draw on its representation as a network graph, where the subunits become nodes and the identified interactions become links. Indeed, this way of summarizing knowledge has become so successful that we witness a rapidly increasing interest in the graph-properties of such network depictions [1–4]. The use of networks as a tool to represent and analyze functional interactions has been gaining importance also in neuroscience [1, 5–9]. In neuroscience, however, it is often overlooked that all derived graph measures are only as good as the reconstruction of the underlying interactions. This reconstruction may suffer, because the identification of all interactions in a multi-node network is fundamentally intractable since it poses a problem in the complexity class of so called "NP-hard" problems [10, 11]. Thus, true network graphs of interactions must be recovered using approximations if we do not want to forgo the use of network representations altogether.

To see why the identification of all interactions in a multi-node network is fundamentally intractable, we have to consider that next to the interactions from one node simply to one other node (a bivariate or pairwise interaction), there may well be interactions from a set of two (or more) source nodes to a target node. Moreover, this multivariate nature of the interactions makes it necessary to control for a parallel influence from any other source in the network when trying to determine whether a particular set of source nodes interacts with the target node in question. It is the enormous number of combinations of potential sources and parallel influences that makes it impossible to search all possibilities in reasonable time for any but the smallest systems (e.g. $n < 20$, [10, 12]). In fact, it can be shown formally that the problem belongs to the class of NP-hard problems, which are believed to lack algorithms that produce solutions for arbitrary input sizes in polynomial time [13].

To nevertheless apply graph theory and network models in neuroscience we need to resort to approximate representations of the true multivariate interactions. Here, the term approximation implies that we will have to commit errors. These errors can be of two types—falsely identifying an interaction that is physically absent, or missing an interaction that is physically present. While both types of errors may have detrimental effects on interpretability of popular graph metrics, we may still ask which type of error to prefer, and how to build fast and efficient approximations that predominantly show the preferred type of error.

Here, we suggest that missing out on interactions instead of including spurious ones may be preferable because the nature of the obtained network becomes more 'reliable' in the sense that all the depicted links do exist. This knowledge can then be built upon in future work. Therefore we present an algorithm that can prune the most frequent spurious interactions from graphs obtained by a simple and efficient bivariate analysis of interactions (this idea was first proposed in [14] in abstract form).

Our focus here is specifically on corrections of graphs obtained from bivariate (i.e., pairwise) analysis methods as these have most often been used to overcome the intractability of the full network reconstruction described above. Despite their popularity, iterative bivariate analyses introduce well known methodological artifacts in the reconstructed interactions [12, 15, 16]: (1) Bivariate analysis may detect *spurious interactions* (false positives) whenever the dependency between two time series is caused or mediated by one or more additional nodes in the network; (2) bivariate analysis may miss *synergistic effects* [17, 18] that two or more time series have on a third. These two problems diminish not only the reliability of individual links in a network but also compromise graph metrics of the global network.

In this study we investigate a solution to the first problem above. Our solution builds on the possibility of reconstructing the delays of interactions (e.g. [19] and similar approaches), and on specific interaction-delay based fingerprints that potentially spurious interactions must leave even in a bivariate analysis. Our method allows to tag potentially spurious interactions for further testing (e.g. by a targeted multivariate analysis) or to remove them entirely from the graph to obtain its most reliable core. Our method thus keeps the advantages of bivariate methods in terms of data efficiency and computational tractability over approaches that are approximately or fully multivariate.

In the following we first provide the necessary background on delay reconstruction by information theoretic methods, and on graphs. Then we present our algorithm and a reference implementation as part of the open source toolbox TRENTOOL ([20], www.trentool.de). Subsequently, we characterize its properties and limitations based on theoretical considerations, simulations and application to magnetoencephalographic (MEG) data. We discuss the relative merits of our approach and other possible approximations to a fully multivariate analysis of networks, and close by outlining possible strategies to deal with the identified potentially spurious links in the network.

## Background and Implementation

Before we outline the algorithm in more detail, we will provide some background information by reviewing the recovery of interaction delays in an information theoretic framework. We will also describe the coupling motifs leading to the detection of spurious interactions. Subsequently, we will formalize the network concept in mathematical terms and complement the common undirected and unweighted network representation used for neural data [1, 21] by introducing the weighting of network connections with their respective interaction delays (using the estimator provided in [19]). We will then describe the rationale underlying the algorithm and its implementation. We conclude this section with the validation of the algorithm using simulated as well as experimental data.

### Background

**Interaction delay reconstruction.** Our algorithm is based on the availability of the interaction delays for bivariately reconstructed interactions. We will systematically use the term "bivariate interaction" to indicate that in the bivariate analysis setting there is no guarantee that a reconstructed interaction is actually present in the underlying data; nevertheless, even for a spurious interaction a meaningful delay can be assigned (see examples in [19]). One possibility to obtain the delays for bivariate interactions is to use delay-sensitive measures of information transfer, i.e. transfer entropy (TE) estimators. In [19] we presented a delay-sensitive TE functional:

$$TE_{SPO}(X \rightarrow Y, t, u) = \sum_{y_t, \mathbf{y}_{t-1}, \mathbf{x}_{t-u}} p(y_t, \mathbf{y}_{t-1}, \mathbf{x}_{t-u}) \log \frac{p(y_t | \mathbf{y}_{t-1}, \mathbf{x}_{t-u})}{p(y_t | \mathbf{y}_{t-1})}, \quad (1)$$

which quantifies the mutual information between the past state $\mathbf{x_{t-u}}$ of a source $X$ and the present value $y_t$ of a target $Y$ at a specific time delay $u$—conditional on the past state of the target, $\mathbf{y_{t-1}}$.

This functional can be used to recover the physical interaction delay $\delta_{X,Y}$ by scanning over possible values for $u$, and by taking the value of $u$ where TE reaches a maximum as the (bivariate) interaction delay:

$$\delta_{X,Y} = \arg\max_u (TE_{SPO}(X \rightarrow Y, t, u)). \quad (2)$$

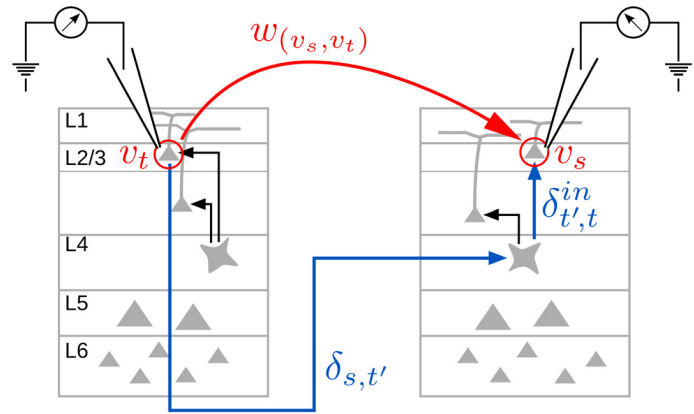

**Fig 1. Reconstructing delay times from electrophysiological recordings.** The physiological delay between two measurement points $v_s$, $v_t$ consists of the time needed for information transfer via axonal connections ($\delta_{s,t'}$) and internal computation within populations of neurons ($\delta_{t',t}^{in}$), such that in a network representation of reconstructed interactions we find $w_{(v_s,v_t)} = \delta_{s,t'} + \delta_{t',t}^{in}$. Black arrows represent information transfer within the neural microcircuits.

A similar approach may be used for Granger causality as TE is equivalent to Granger causality for data with a jointly Gaussian distribution [22].

Note, that reconstructed interaction delays incorporate not only the mere transfer time between two neural processes, but also the time needed for local computation, if we only obtain one channel per subunit (see Fig 1 for further explanation). Thus, interaction delay reconstruction as proposed in [19], captures the total delay between two measurement *points*, which in a neural system may consist of transfer time along an axonal connection but also of time needed for information transfer within the local neural microcircuit.

**Spurious interactions in bivariate analysis of multivariate data sets.** Spurious interactions may arise in bivariate analysis from one of two distinct coupling motifs: In the first coupling motif (Fig 2A) the dynamics of two or more nodes, representing neural sources, are simultaneously driven by processes in a third node. A bivariate analysis may detect an interaction between the two driven nodes. We term this a *common drive* (CD, also "common cause" [12]). In the second coupling motif (Fig 2B) an interaction between two nodes is mediated by one or more intermediate nodes in the network and information transfer from source node to target node is routed via these intermediate nodes. A bivariate analysis may detect an interaction between the source and target node. We term this a *cascade effect* (CE, also "pathway effect" [12] or "indirect causal pathways" [15]). The detection of spurious interactions by bivariate analysis have been demonstrated for simulated data as well as in neural recordings [15, 16, 23–25].

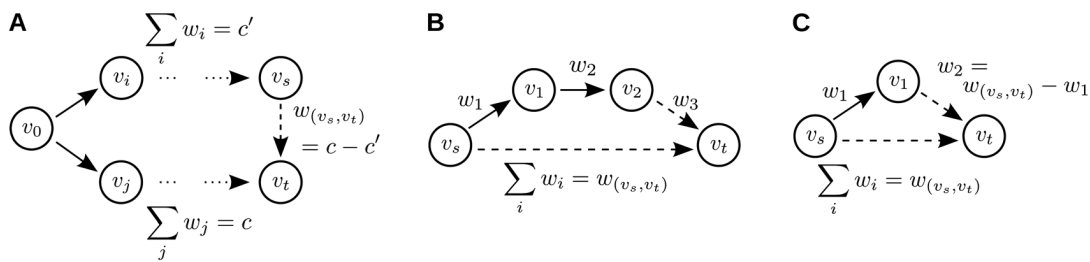

**Fig 2. Spurious Interactions.** (A) Common drive effect: A spurious interaction due to common drive may be potentially present if the processes at vertices $v_s$ and $v_t$ are driven by $v_0$ with differential delays, such that the bivariate information transfer between $v_s$ and $v_t$ is a result of the common input from $v_0$; (B) Cascade effect: Spurious interaction due to cascade effects may be potentially present for all cascades of information transfer in a "chain" of sources. In the example here, the bivariate information transfer between $v_s$ and $v_t$ (edge $(s, t)$) can be explained by an alternative routing of information via vertices $v_1$ and $v_2$. The summed weight of the alternative routing is equal to $w_{(s,t)}$; (C) "Triangle" motif: This is the most simple motif potentially giving rise to either of the above spurious interactions: $(v_s, v_t)$ could be a result of a cascade effect with respect to the path $\langle v_s, v_1, v_t \rangle$, and $(v_1, v_t)$ could be the result of a common drive of $v_1$ and $v_t$ by $v_s$. At most one of these interactions can be spurious.

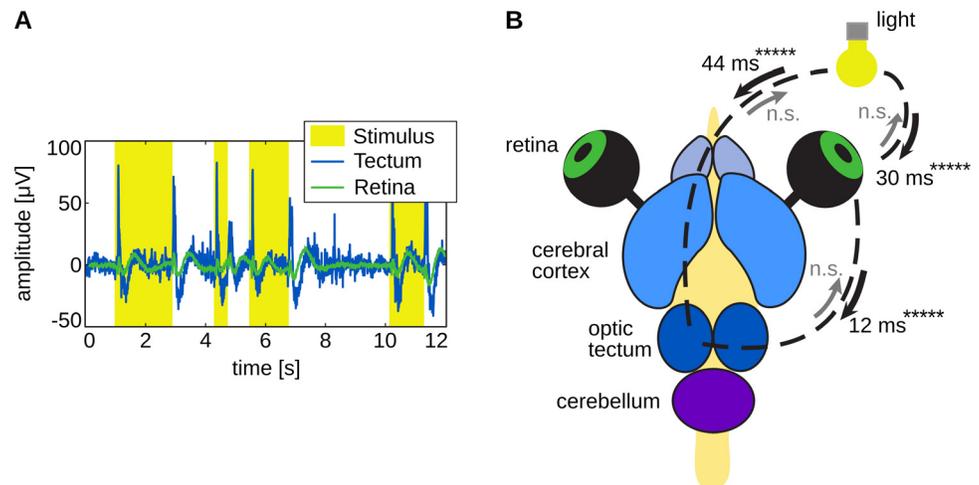

**Fig 3. Directed interactions in the turtle brain during visual stimulation with random light pulses (modified from [19], creative common attribution license CC BY).** (A) Raw traces recorded in the tectum (blue) and from the retina (green) overlaid on the light pulses (yellow). (B) Turtle brain explant with eyes attached. Transfer entropy was found from the retina of the right eye to the left tectum, as well as from the light source (yellow) to the retina and to the tectum (***** denotes $p < 10^{(-5)}$). P-values for the opposite directions were not significant (*n.s.*). Note, that the interaction between light source and optic tectum shows a interaction delay roughly equal to the summed interaction delay between light source and retina and retina and optic tectum (deviation $\leq 5\%$).

Coupling motifs leading to spurious interactions due to CD and CE exhibit a specific *timing signature* in the network of bivariately reconstructed interactions. We found an example for such a timing signature in experimental data recorded from the turtle (Fig 3C) [19]). Here, a spurious interaction was detected between the light source and the optic tectum. This spurious interactions resulted from a CE, i.e., an actual routing of information from light source to tectum via an intermediate node, namely the retina. Information transfer delays reconstructed with the TE estimator proposed in [19], revealed this CE: The summed interaction delays in the actual routing of information equaled the delay of the spurious information transfer.

**Graph representation of neural data and notation.** As a last preliminary, we will present the mathematical formalization of a network to give a precise account of the algorithm and its functionality in the subsequent section. Table 1 lists the most important variables. In mathematical terms, a network is described by a (directed) graph $\mathbf{G} = \{\mathbf{V}, \mathbf{E}\}$, where $\mathbf{V}$ denotes a set of vertices or nodes and $\mathbf{E} \subseteq \mathbf{V} \times \mathbf{V}$ represents a set of connections between nodes, called edges [26]. In a neuroscience application, $\mathbf{V}$ may represent a set of individual functional units $v_i$, e.g. neurons, sources in MEG analysis, or voxels in functional magnetic resonance imaging data; $\mathbf{E}$ may represent some sort of connection between two units, for example significant functional interactions. An edge is written as a tuple $(v_i, v_j)$, representing an edge between any two sources $v_i$ and $v_j$. Note, that such a tuple $(v_i, v_j)$ defines an edge as an ordered pair of two vertices and as such indicates a *directed* connection between these two sources (the two elements of the tuple are not interchangeable, such that $(v_i, v_j) \neq (v_j, v_i)$). We further assume that edges are weighted by a weighting function $w : E \mapsto \mathbb{N}$ that maps the set of edges to the natural numbers. Here, these natural numbers are chosen proportionally to the timing information as precisely and as parsimoniously as possible. We call $w_{(v_i, vj)}$ the weight of edge $(v_i, v_j)$.

**Table 1. Notation.**

| Graph representation | |
|---|---|
| $G$ | graph, consisting of the sets $V$ and $E$ |
| $V$ | set of vertices |
| $E$ | set of edges |
| $v_i$ | vertex with index $i$ |
| $(v_i, v_j)$ | edge from $v_i$ to $v_j$ |
| $w_{(v_i, vj)}$ | weight of edge $(v_i, v_j)$ |
| $v_i \rightarrow v_j$ | path from $v_i$ to $v_j$ |
| $v_i \overset{k}{\rightarrow} v_j$ | path from $v_i$ to $v_j$ with summed weight $k$ |
| $l$ | path length, i.e. no. edges in a path |
| **Algorithm** | |
| $(v_a, v_b)$ | edge under investigation in current algorithmic iteration |
| $w_{(v_a, vb)}$ | weight of the edge under investigation |
| $\theta$ | threshold to account for imprecisions in interaction delay reconstruction |
| $w_{crit}$ | critical path weight, $w_{crit} = w_{(v_a, vb)} + \theta$ (target weight of the algorithm) |
| $v_s, v_t$ | start and target node ($v_s = v_a$, $v_t = v_b$) |
| $L^n_{v_t}(w_i; \rightarrow v_j)$ | set of solutions in algorithmic step $n$, that solve the subproblem $v_s \overset{w_i}{\rightarrow} v_j$ |

For any edge $(v_i, v_j)$, $v_i$ is considered the *predecessor* of $v_j$. $v_j$ is called the *child* of $v_i$. A *path* $v_0 \rightarrow v_l$ is defined as a sequence of vertices $\langle v_0, v_1, \ldots, v_i, \ldots, v_{l-1}, v_l \rangle$, where every two consecutive vertices $v_i\, v_{i+1}$ are connected by an edge $(v_i, v_{i+1})$. We call $l$ the length (number of edges) of the path and we will refer to this length $l$ as a path's *graphical length*, describing the number of edges used to graphically represent the path in the graph. The total weight of a path is the sum of the weights of all individual edges comprising the path, $\sum_i w_{(v_i, vi+1)}$.

[Fig 4](#) gives a schematic overview of the construction of a graph from time series data recorded from a set of neural sources. Edges in the graph represent significant interactions between sources (vertices); edge weights represent reconstructed interaction delays. We here use TE to analyze directed interactions, using the estimator proposed in [19] to recover significant TE and corresponding interaction delays, but other approaches are possible.

## Rationale and implementation of the algorithmic solution

**Rationale of the algorithm.** Based on the graph representation of reconstructed directed interactions, their delays, and the theoretical preliminaries presented in the last subsection, we will now propose an algorithm that detects potentially spurious interactions by exploiting the timing signature of CE and simple CD.

As input the algorithm expects a directed, delay-weighted graph $G := \{V, E\}$, represented by its connectivity matrix. Connections are weighted with the estimated interaction delays $w_{(v_a, v_b)} = \hat{\delta}_{(v_a, v_b)}$, i.e., the estimated physical delays between the processes represented by $v_a$ and $v_b$. Note that such a graph needs to be constructed from a connectivity measure, which is (a) directed and (b) allows for the reconstruction of interaction delays. Additionally, the user has to provide a threshold $\theta$ to account for noise in empirical measurements as well as imprecision in analysis methods (described in detail below). We furthermore assume that weights have been linearly scaled, such that they do not have any decimal places and can be represented by integer values.

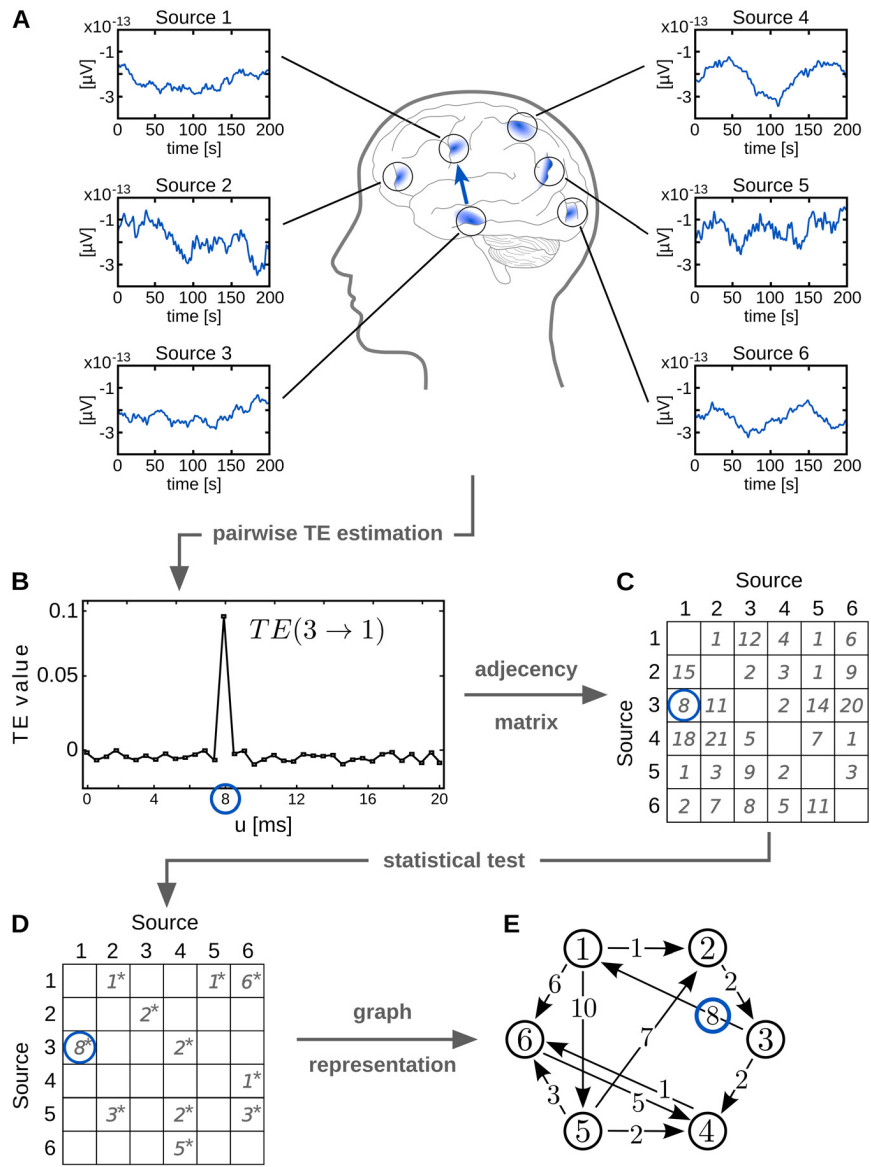

**Fig 4. Graph representation of neural data.** (A) Recorded signals from various sources in the brain; (B) Pairwise estimation of transfer entropy (TE) and reconstruction of interaction delays $u$ between any two sources; (C) Adjacency matrix: representation of estimated delay times between all source combinations, every entry represents an information transfer from the $i$th row to the $j$th column; (D) Adjacency matrix after test for statistical significance; (E) Visualization of the graph represented by the connectivity matrix: every source is represented by a vertex, every significant information transfer is represented by an edge. (The blue circle indicates the respective representation of an exemplary interaction between source 1 and source 3 throughout all steps of graph reconstruction.)

As a first step, we identify potential CEs by assuming that if a CE is present in the data, the bivariate interaction represented by any edge $(v_a, v_b) \in E$ with weight $w_{(v_a, v_b)}$ can be explained by an alternative routing of information via intermediate vertices (see an example in Fig 2B). Thus, iteratively for every edge $(v_a, v_b)$ in **E** the algorithm sets $v_a = v_s$ as the starting and $v_b = v_t$ as the target node of the current iteration. Then the algorithm searches for an alternative path for $(v_s, v_t)$, where a path is assumed to be an alternative path if the summed delay interaction times $\sum_i w_{(v_i, vi+1)}$ of all edges in the path are (approximatively) equal to $w_{(v_s, vt)}$, i.e., $w_{(v_s, vt)} - \theta < \sum_i w_{(v_i, vi+1)} < w_{(v_s, vt)} + \theta$. For an example see Fig 2B, where edge $(v_s, v_t)$ (dashed arrow) has a graphically longer, alternative path $\langle v_s, v_1, v_2, v_t \rangle$ with equal summed weight (solid arrows). If the algorithm finds such an alternative path, the edge currently under investigation is tagged as potentially spurious.

In a second step, the algorithm additionally identifies potential simple CD effects based on the results from the first analysis step. Simple CD effects occur in graph motifs that form acyclic "triangles" (Fig 2C). We define an acyclic triangle as any three nodes that are acyclic, pairwise connected. We suspect a spurious link due to simple CD if a triangle motif exhibits a suspicious timing signature; these suspicious triangles are identified from the results of the first analysis step by listing all edges with alternative paths of graphical length two. We propose to tag two edges in each of these identified triangles: (1) the direct edge due to a CE (edge $(v_s, v_t)$ in Fig 2C) and (2) the second edge due to a simple CD (edge $(v_1, v_t)$ in Fig 2C). Note however, that of both tagged edges one has to be a non-spurious interaction for this rationale to hold: if the CE link $(v_s, v_t)$ is rejected, a possible driver-target relationship between nodes $v_s$ and $v_t$ is destroyed, thus nullifying the argument for the simultaneous rejection of edge $(v_1, v_t)$. The same argument holds vice versa: a rejection of the tagged edge $(v_1, v_t)$ due to CD destroys the information cascade from $v_s$ to $v_t$ and thus cancels the CE causing the detection of $(v_s, v_t)$. Thus, tagging both edges in a triangle motif yields a slightly too conservative approach to network representation. For the treatment of tagged links in neuroscience we refer to the *Discussion* section.

A further important consideration is that once the algorithm has detected an alternative path $v_a \rightsquigarrow v_b$ for an edge $(v_a, v_b)$, this alternative path stays intact, even if, at a later step, some edge in $v_a \rightsquigarrow v_b$ is tagged (and probably removed). Assume, that we have an edge $(v_a, v_b)$ with an alternative path $v_a \rightsquigarrow v_b$ and that at a later step, one or more edges in this path get tagged. Then, for each of these tagged edges, leaving a 'gap' in $v_a \rightsquigarrow v_b$, an alternative path of equal summed weight exists by definition of the algorithm. This alternative path closes the "gap" in $v_a \rightsquigarrow v_b$, such that nodes $v_a$ and $v_b$ are still connected by a new path $v_{a'} \rightsquigarrow v_{b'}$. This new path has the same summed delay as $v_a \rightsquigarrow v_b$, but is graphically longer. Therefore, alternative paths identified at some step in the algorithm remain intact even if links within the paths are tagged at a later point in the algorithm's execution.

**Implementation overview.** We now present the implementation of the strategy described above; more precisely, we present an algorithm that finds all alternative paths for any given edge $(v_a, v_b) \in \mathbf{E}$, given a directed, weighted graph $\mathbf{G} = \{\mathbf{V}, \mathbf{E}\}$. See Fig 5 for an overview of the algorithm.

As a preprocessing step, we construct a graph $\mathbf{G}'$ by removing the edge $(v_a, v_b)$ from $\mathbf{G}$ and relabeling nodes $v_a$, $v_b$ as starting and target nodes $v_s$, $v_t$ of the current iteration of the algorithm. The target weight of the alternative path is set to $w_{crit} = w_{(v_a, vb)} + \theta$. Furthermore, $\mathbf{G}'$ is represented as an inverted adjacency list, i.e., a list of all nodes in $\mathbf{V}$, where for every node all of its predecessors are listed. $v_s$ is set as the first node in this list (note that we assume that $v_s$ has index 1), $v_t$ is set as the last node (has index $|\mathbf{V}|$). Therefore, for all other nodes $v_j$ in the adjacency list it now holds that $2 \leq j \leq |\mathbf{V}|$.

After preprocessing the input, alternative paths $v_s \rightsquigarrow v_t$ with weight $w_{(v_s, vt)} \pm \theta$ are detected in two steps: (1) a memoized dynamic programming approach [27] is used to determine, whether any path $v_s \rightsquigarrow v_t$ of a total weight $w_{(v_s, vt)} \pm \theta$ exists; (2) a modified depth first search (DFS, see [27] for a description) is used to reconstruct all paths with weights in the interval $w_{(v_s, vt)} \pm \theta$ from the solution obtained in step (1). The second step is necessary to reject paths that contain loops and to allow for further analysis (for example the identification of triangle motifs).

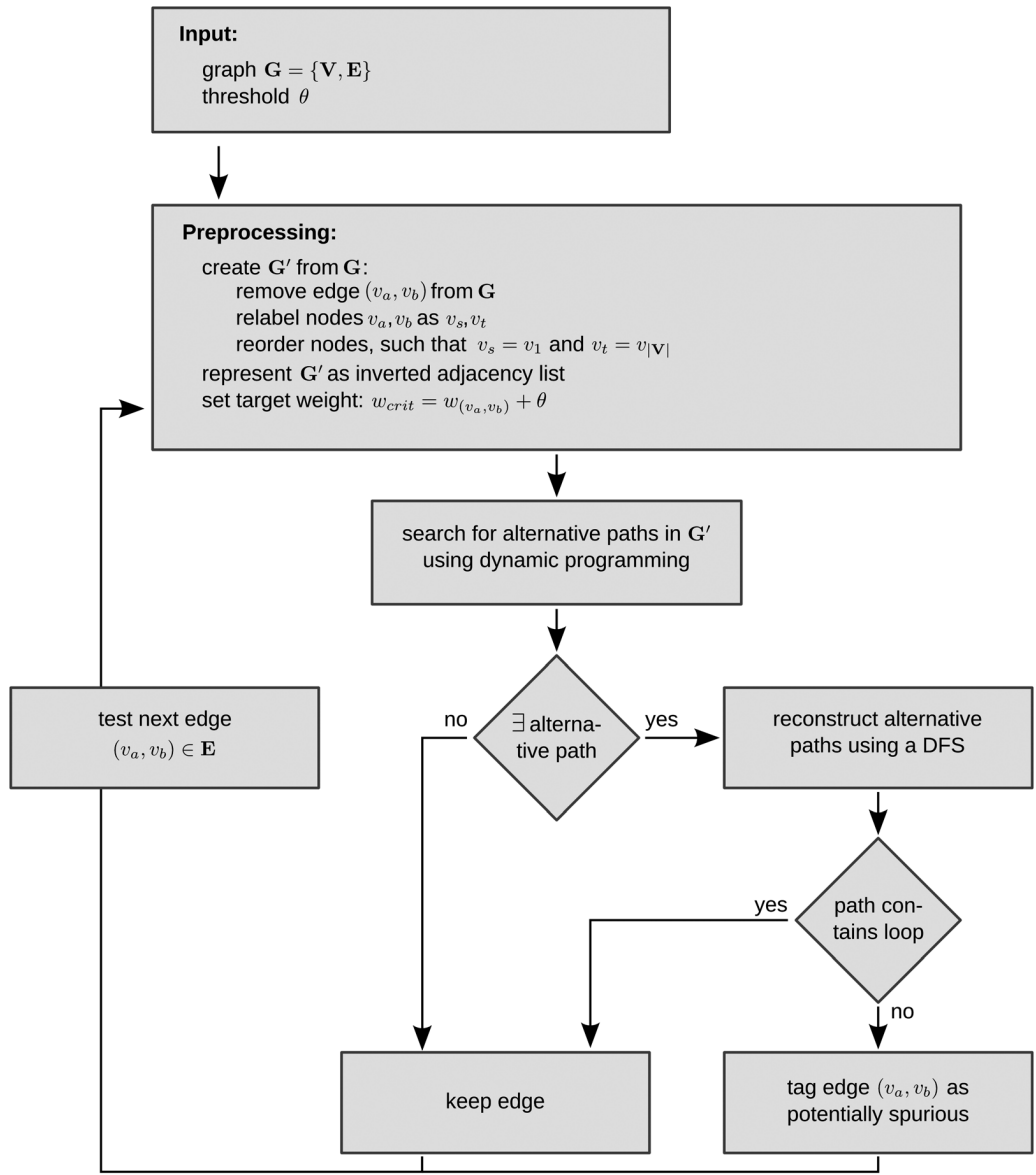

**Fig 5. Overview of the proposed algorithm.** The algorithm expects a weighted and directed graph $\mathbf{G} = \{\mathbf{V}, \mathbf{E}\}$ and a threshold $\theta$ as input. In a preprocessing step, the algorithm creates graph $\mathbf{G}'$ from input $\mathbf{G}$, as an input for the dynamic programming algorithm, by removing edge $(v_a, v_b)$ and by relabeling and reordering nodes. Then, in the next step, alternative paths for $(v_a, v_b)$ are searched through dynamic programming (see also main text). If at least one alternative path is found, paths are reconstructed using a depth first search (DFS, [27]) to ensure that alternative paths do not contain loops. If an alternative path contains no loops, the currently investigated edge $(v_a, v_b)$ is tagged as potentially spurious. If no alternative edge is found, $(v_a, v_b)$ is considered non-spurious. The algorithm then enters the next iteration, in which the next edge $(v_a, v_b) \in \mathbf{E}$ is investigated for alternative paths.

doi:10.1371/journal.pone.0140530.g005

The algorithm was implemented as part of the open source toolbox TRENTOOL [20].

**Dynamic programming.** We use a dynamic programming approach in step (1) to handle the inherent complexity of the problem at hand (see *Discussion*). Dynamic programming allows for the solution of a complex problem by decomposing it into easily solvable subproblems. By starting with trivial base cases, subproblems of increasing complexity are solved iteratively by taking recourse to tabulated solutions of previous (more simple) subproblems. This reduces computational demand and is repeated until the algorithm reaches the most complex subproblem, which represents the input problem. In the following, we will first define the subproblems to the present problem, and then describe the algorithmic solution to an individual subproblem. A detailed, graphical account of both steps is presented in [Fig 6].

**Formulation and ordering of subproblems.** The overall problem of finding a path from $v_s$ to $v_t$ with weight $w_{(v_s, \, vt)} \pm \theta$ is divided into subproblems by asking, whether a "simpler" path $v_s \xrightarrow{w_i} v_j$ exists; $v_j$ is any node in $\mathbf{V}$ and $w_i$ is any path weight $w_i \leq w_{crit} = w_{(v_s, \, vt)} + \theta$. For example, finding a path $v_1 \xrightarrow{3} v_3$ is a subproblem to finding path $v_1 \xrightarrow{9} v_{10}$ where $v_1 = v_s$ and $v_{10} = v_t$.

The presented algorithm solves subproblems iteratively for increasing complexity using a dynamic programming approach. It is thus necessary to order the subproblems by their complexity and to make their solutions immediately accessible for reuse in subsequent algorithmic steps. This is realized by organizing solutions in a two-dimensional *solution array*, in which solutions are ordered by path weights and node indices, the two parameters determining complexity. Starting with the most simple base case where $w_i = 0$ and $v_s = v_1 = v_j$ (this subproblem describes a path $v_s \xrightarrow{0} v_s$), subsequent subproblems are formulated by increasing path weights and node indices in integer steps. Thus subproblems are formulated for all combinations $w_i = 1, 2, \ldots, w_{crit}$ and $v_j \in \mathbf{V}$ (see for example [Fig 6C] for the first iteration of path weights and a complete iteration over all vertices). Individual solutions to subproblems are tabulated in the solution array of size $[0; w_{crit}]$ by $[1; |\mathbf{V}|]$ and are indexed by the current values for $w_i$ and $v_j$. This organization of subproblems allows for the easy retrieval of solutions from earlier iterations to solve the subproblem currently at hand (see below and [Fig 6]).

**Finding solutions to subproblems.** For any given subproblem $v_s \xrightarrow{w_i} v_j$ the algorithm determines whether a path of weight $w_i$ leads into node $v_j$. The algorithm does this by testing whether any single edge leading into $v_j$ together with the solution to a simpler subproblem forms the path $v_s \xrightarrow{w_i} v_j$. In particular, for every edge $(v_p, v_j)$ leading into $v_j$ (where $v_p$ is a predecessor of $v_j$), the algorithm checks if this edges extends a path leading into $v_p$ such that the resulting path solves the current subproblem $v_s \xrightarrow{w_i} v_j$ (see [Fig 7A]). We call the treatment of one edge $(v_p, v_j)$ an *algorithmic step*. In one algorithmic step, the algorithm checks (1) if there *exists* a path to $v_p$ and if so, (2) whether the path has length $w_p = w_i - w_{(v_p, \, vj)}$. If both conditions are met, the currently considered edge $(v_p, v_j)$ together with the path to predecessor $v_p$ solves the current subproblem $v_s \xrightarrow{w_i} v_j$ (because a path $v_s \xrightarrow{w_p} v_p$ exists, that together with $(v_p, v_j)$ forms a path of summed weight $w_i = w_p + w_{(v_p, \, vj)}$). The algorithm terminates once the most complex subproblem has been investigated, i.e., it has been tested if a path of length $w_{crit}$, connecting node $v_s$ to $v_t$ can be found.

Note that "checking" if a path of weight $w_p$ to node $v_p$ exists corresponds to looking up whether a solution to the subproblem $v_s \xrightarrow{w_p} v_p$ exists in the solution array, i.e., the algorithm has to look up the entry in the solution array at row $p$ (for node $v_p$) and column $w_i - w_{(v_p, \, vj)}$ (for weight $w_p$). In doing so, the algorithm solves the subproblem by reusing earlier solutions. Note also, that relevant subproblems are guaranteed to have been solved at an earlier point in the execution of the algorithm as the algorithm treats subproblems in the order of increasing complexity: subproblems are solved by first iterating over all path weights $w_i = 1, 2, \ldots, w_{crit}$ in an outer loop, and second iterating over all nodes from $v_s := v_1$ to $v_t := v_{|V|}$ in an inner loop; a relevant subproblem $v_s \xrightarrow{w_p} v_p$ is guaranteed to have been solved because it always holds that $w_p < w_i$ per definition of $w_p$.

To tabulate solutions to subproblems, for every edge solving the subproblem and its weight, we add a tuple $(w_{(v_p, \, vj)}, v_p)$ to a set of solutions. More specifically, when iterating over all $N$ potential incoming edges $(v_p, v_j)$, we enter valid solutions in a sequence of sets indexed by the current algorithmic step $n$: $\mathbf{L}^0_{w_i, v_j} \subseteq \ldots \subseteq \mathbf{L}^n_{w_i, v_j} \subseteq \mathbf{L}^{n+1}_{w_i, v_j} \subseteq \ldots \subseteq \mathbf{L}^N_{w_i, v_j}$. Each tuple in $\mathbf{L}^N_{w_i, v_j}$ then indicates the existence of one alternative path for the subproblem. The collection later allows the reconstruction of all alternative paths for the overall problem.

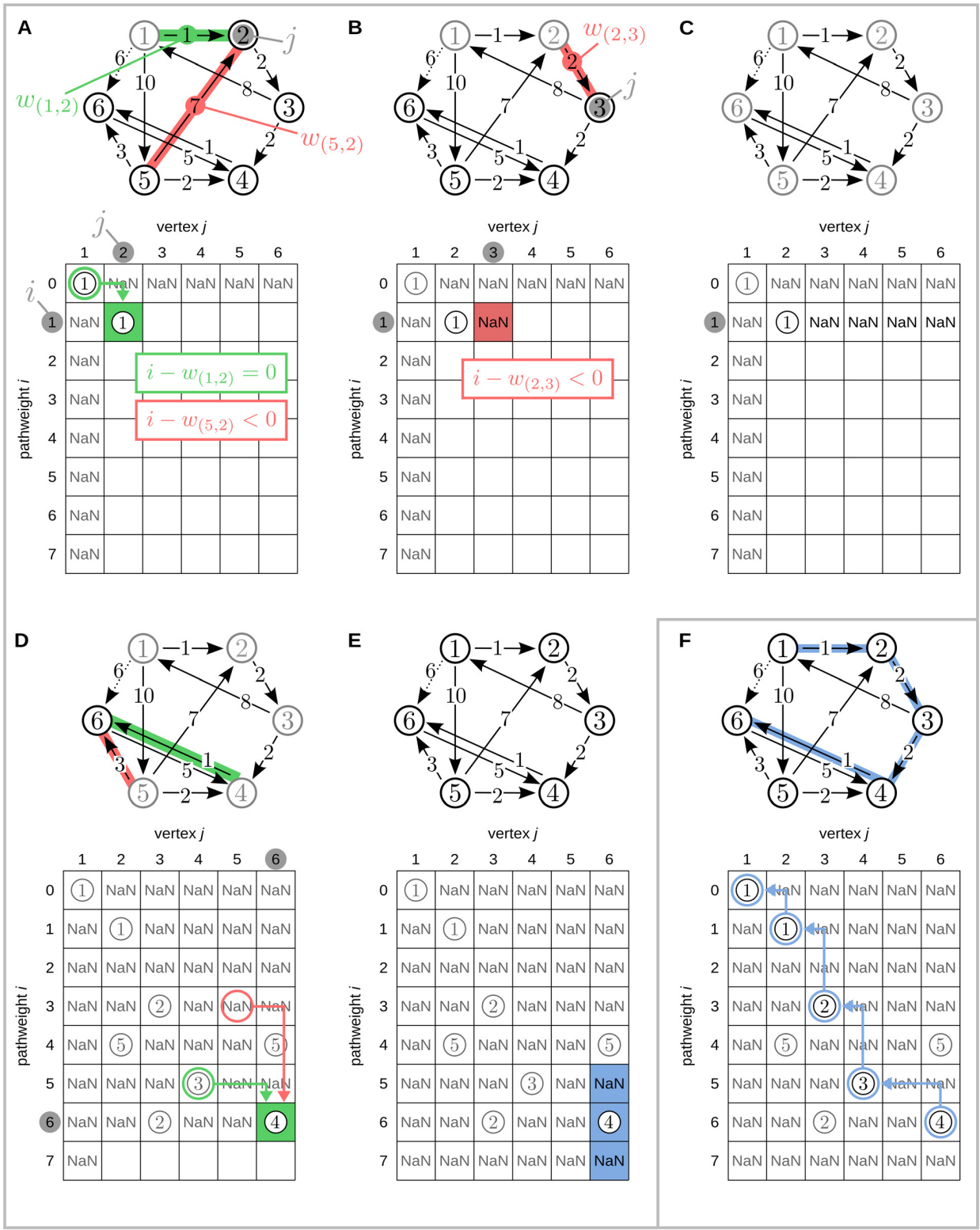

**Fig 6. Visualization of the proposed algorithm.** Search for alternative paths to edge (1, 6) (dotted arrow), i.e., $v_s = 1$ and $v_t = 6$. Solutions $\mathbf{L}_{v_s}^n (w_i; \neg v_j)$ to subproblems are managed in a two dimensional solution array indexed by path weight $w_i$ and vertex number $v_j$. Solutions are calculated iteratively over $w_i$ (rows) and $v_j$ (columns). (A) Solution matrix after first iterative step (subproblem $\mathbf{L}_{v_s}^j (1; \neg v_2)$): There are two edges leading to vertex 2 of which only edge (1, 2) yields a valid solution by pointing to an earlier solved subproblem (green box), whereas edge (5, 2) has a weight of 7 leading to a negative difference in weights $i - w_{(5,2)}$ (red box) for which no earlier solution exists; (B) Solution matrix after third iterative step $\mathbf{L}_{v_s}^3(1; \rightsquigarrow v_3)$: Here, no valid solution exists (none of the arrows leading to vertex 2 are part of a path with summed weight 1); (C) Solution array after iteration over all vertices $v_j$ for $w_i = 1$ (all vertices have been checked for a path of weight 1, originating from the start vertex 1); (D) Solution to subproblem $\mathbf{L}_{v_s}^n(6; \rightsquigarrow v_6)$: edge (4, 6) together with the solution $\mathbf{L}_{v_s}(5; \rightsquigarrow v_1)$ form a valid path, whereas edge (5, 6) is not part of a valid solution as $\mathbf{L}_{v_s}(3; \rightsquigarrow v_5)$ is empty; (E) The algorithm terminates after iteration over all vertices $v_j$ and path weights up to $w_{(v_s, vt)} + \theta$, where $\theta$ is a user defined threshold of 1. Backtracking is conducted for all entries in the reconstruction interval $w_{(v_s, vt)} \pm \theta$ (entries marked blue); (F) Reconstructed alternative path by backtracking of subproblems.

Formally, after initially setting $\mathbf{L}_{w_i, v_j}^0 \equiv \emptyset$, we can define each algorithmic step $n$ as

*For every edge $(v_p, v_j)$ leading into $v_j$:*

$$\mathbf{L}_{w_i, v_j}^{n+1} := \begin{cases} \mathbf{L}_{w_i, v_j}^n \cup (w_{(v_p, v_j)}, v_p) & \text{if } v_j \neq v_s \wedge \mathbf{L}_{w_p, v_p}^M \neq \emptyset \\ \mathbf{L}_{w_i, v_j}^n & \text{if } \mathbf{L}_{w_p, v_p}^M = \emptyset \\ (0, v_s) & \text{if } v_j = v_s. \end{cases} \quad (3)$$

Here, $\mathbf{L}_{w_i, v_j}^n$ denotes the current set of tuples contributing to the solution of the subproblem $v_s \overset{w_i}{\rightsquigarrow} v_j$, i.e., a path leading from $v_s$ to $v_j$, that has a summed weight of $w_i$ in algorithmic step $n$. $\mathbf{L}_{w_i, v_j}^M$ is a set of solutions to the subproblem $v_s \overset{w_p}{\rightsquigarrow} v_p$ investigated earlier (where solutions were collected over $M$ algorithmic steps). Formula 3 expressed that for every edge $(v_p, v_j)$, it is tested if two conditions are met (see also Fig 7B):

1. there exists a solution to the previous subproblem $\mathbf{L}_{w_p, v_p}^M$, i.e., a path to the predecessor $v_p$ with weight $w_p = w_i - w_{(v_p, vj)}$;

2. the edge $(v_p, v_j)$ from predecessor to current node has a weight such that $w_p + w_{(v_p, vj)} = w_i$.

If both conditions are met, then the tuple $(w_{(v_p, vj)}, v_p)$ is added to $\mathbf{L}_{w_i, v_j}^n$. When all edges $(v_p, v_j)$ have been tested, the next subproblem $v_s \overset{w_i}{\rightsquigarrow} v_{j+1}$ (inner loop) or $v_s \overset{w_{i+1}}{\rightsquigarrow} v_j$ (outer loop) is considered. The algorithm terminates once all subproblems have been investigated.

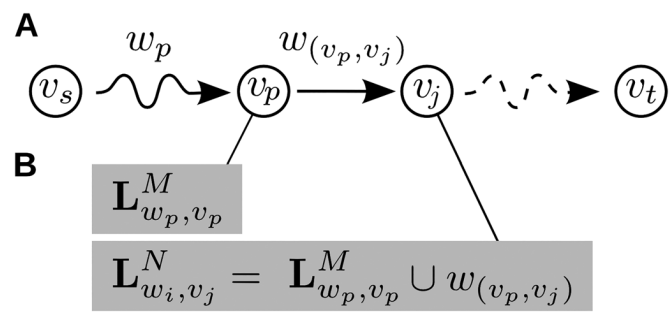

**Fig 7. Schematic example of a subproblem of the proposed algorithm.** (A) Example subproblem $\mathbf{L}_{w_i, v_j}^n$: At the $n^{th}$ algorithmic step, we search for all paths of weight $w_i$ leading to node $v_j$; (B) Finding a solution for the current subproblem by investigating solutions to prior subproblems: We investigate all predecessors $v_p$ of the current node $v_j$; if there exists a solution to $\mathbf{L}_{w_p, v_p}^M$, i.e., there is a solution to the prior subproblem $v_s \overset{w_p}{\rightsquigarrow} v_p$ of finding a path of weight $w_p$ leading from $v_s$ to $v_p$, and $w_p + w_{(v_p, vj)} = w_i$, we find a solution to the current subproblem $\mathbf{L}_{w_i, v_j}^n$.

**Backtracking.** In a second step, the algorithm uses backtracking to reconstruct relevant paths from the solution array returned by the dynamic programming step described in the last subsection (see Fig 6F). Paths are considered relevant, if they lead to the current target node $v_t$ and have a summed weight of $w_{(v_s,\ vt)} \pm \theta$. Thus, paths are reconstructed from all entries that correspond to the solutions to subproblems $v_s \rightsquigarrow v_t$ with weight $w_{(v_s,\ vt)} \pm \theta$ (we call these relevant entries in the solution array *reconstruction interval*).

The backtracking algorithm uses depth first search (DFS, for a more detailed description see for example [27]) to reconstruct all paths starting from one entry in the reconstruction interval at a time; it is therefore called for each entry in the reconstruction interval individually (for example, in Fig 6F, this corresponds to three calls of the backtracking algorithm for fields [5, 6], [6, 6] and [7, 6]). The backtracking is done by recursively expanding each entry in the currently considered field (i.e., visiting the next field, indicated by the currently considered solution to a subproblem). For example, in Fig 6F, the field [6, 6] points to the field [5, 4], which points to [3, 3] and so on. While expanding one path, the algorithm checks, whether a node in the currently reconstructed path has already been visited during the recursion. If this is the case, the path contains a loop, i.e., a node is visited twice in one path, and the respective path is discarded as it is not a valid solution to the overall problem.

All remaining reconstructed paths are considered alternative paths to the edge $(v_s, v_t)$. If at least one such alternative path exists, $(v_s, v_t)$ is considered potentially spurious due to a CE.

**Additional analysis of triangle motifs.** As laid out in the subsection *Rationale of the algorithm*, the algorithm identifies simple CD in an additional step. Simple CD occurs in triangle motifs, which can be identified by listing all edges with valid alternative paths of length two. In each triangle, the second edge of the alternative path is considered as potentially spurious due to a simple CD effect additionally to the edge considered spurious due to CE (see coupling motifs shown in Fig 2C).

**Output.** The algorithm returns a list of potentially spurious edges in **E**, and tags these spurious edges as a CE (identified by an alternative path) or a simple common drive effect (identified in a triangle motif).

## Evaluation

To test the proposed algorithm for correctness and performance in terms of execution times, we simulated networks of different sizes and densities to serve as input graphs. To further demonstrate the algorithm's applicability to neuroscience data, we applied it to networks derived from electrophysiological recordings during a face recognition task.

### Performance of the algorithm in simulated networks

We simulated networks of different types: small-world networks [28], scale-free networks [29] and random networks of different densities [30, 31]. We chose these topologies because small-worldness and scale-freeness have frequently been reported to occur in functional and anatomical networks derived from neuroscience experiments [21, 32, 33] (for a review see also [7] and [34]).

The performance of the proposed algorithm on the simulated networks was tested by (1) varying the size |**V**| of simulated networks and (2) varying the critical path weight for alternative paths $w_{crit}$. Higher values for both parameters increased computational demand by either increasing input size directly (higher values for |**V**|) or by increasing the likelihood for the detection of an alternative path (higher values for $w_{crit}$).

The dependency of the likelihood of detecting an alternative path on $w_{crit}$ is especially relevant in random network topologies: here, the number of possible alternative paths increases exponentially in $w_{crit}$, as higher values for $w_{crit}$ (relative to individual path weights) increase the number of possible combinations of edges that form a path of weight $w_{crit}$. More precisely, given a random graph $\mathbf{G} = \{\mathbf{V}, \mathbf{E}\}$ and $w_{crit} > |\mathbf{V}|$, the probability for the existence of an edge is $p(e) = \rho = \frac{|\mathbf{E}|}{|\mathbf{V}|(|\mathbf{V}|-1)}$ (where $\rho$ is the density of $\mathbf{G}$); furthermore, the probability of the edge having weight $w$ is uniformly distributed with $p(w) = \frac{1}{w_{max}}$. The number of possible alternative paths can then be calculated as

$$\sum_{w'=1}^{w_{crit}} \left[ \sum_{j=1}^{w'} \left( \binom{w'-1}{j-1} \cdot p(w)^j p(e)^j \right) \right], \tag{4}$$

where $\sum_{j=1}^{w'} \binom{w'-1}{j-1}$ is the number of compositions, i.e., the number of ordered sequences of integers that sum to $w'$, the currently considered summed edge weight. The inner sum is dominated by the growth of the number of compositions given by $\sum_{j=1}^{i} \binom{i-1}{j-1} = 2^{i-1}$. Thus, the number of alternative paths grows in $\Omega(2^{w_{crit}})$, i.e., it grows exponentially in the critical path weight given a sufficiently dense, random network topology.

**Small-world networks.** For the simulation of small-world networks we modified the rule for network generation proposed by Watts and Strogatz in [28]. The network generation was done in two steps with parameters $|\mathbf{V}|$ (number of nodes), $n$ (neighborhood coefficient) and $p$ (rewiring probability):

1. Construct a regular ring lattice with $\mathbf{V}$ nodes, where every node $v_i \in \mathbf{V}$ is connected to its $n$ nearest neighbors $v_j$ (such that $(v_i, v_j) \in \mathbf{E}$). Given that $\mathbf{V} = v_1, \ldots, v_{|\mathbf{V}|}$, each node $v_i$ is connected to its neighbors $v_{i+1}, \ldots, v_{i+n/2}$ and $v_{i-1}, \ldots, v_{i-n/2}$.

2. For every node $v_i$ all edges $(v_i, v_j)$ are rewired with probability $p$ by replacing $(v_i, v_j)$ with $(v_i, v_k)$ where $n$ is chosen randomly with uniform probability from $\mathbf{V} \setminus v_i$ while avoiding loops and multiple edges.

3. Every edge is weighted by a random weight $w_{(v_i, v_j)}$ drawn from an interval $[1; w_{max}]$ with uniform probability

Note, that we made two extensions to the original generation rule proposed in [28]: we simulated network edges as directed and weighted, while the original work by Watts and Strogatz assumes undirected and unweighted edges. We defined connections from every node $v_i$ to its $n$ nearest neighbors as directed and weighted them with values randomly drawn from an interval $[1; w_{max}]$. We set $w_{max}$ to the maximum interaction delay found in an MEG data set also used as a second test case described below. Thus, strictly speaking, only the undirected and unweighted network underlying our test cases had small-world properties (i.e., a high clustering coefficient and a short characteristic path length). We used this approximation of small-world properties in a weighted and directed network, as there is no agreement over how directedness and edge weights are to be incorporated into the original notion of small-worldness (as both parameters may alter the global behavior commonly observed in undirected and unweighted small-world networks) [35].

**Scale-free networks.** Scale-free networks were simulated using an algorithm proposed in [36], following an implementation of the rationale by Barabási and Albert [29] in [37] (see also [38]).

Scale-free networks resemble small-world networks in their topology, i.e., they exhibit high local clustering and low characteristic path lengths. Both network types differ however in their degree distributions $p(n)$, the probability that a node interacts with $n$ other nodes in the network: in small-world networks the degrees are normally distributed, while in scale-free networks the degrees follow a power law $p(n) \sim n^{-\gamma}$ (where $\gamma$ may vary for different networks) [29].

**Random networks.** We created random networks of size $|\mathbf{V}|$ by independently including weighted edges with probability $\rho$[31], where $\rho$ denotes the density or edge probability of a graph:

$$\rho = \frac{|\mathbf{E}|}{|\mathbf{V}|(|\mathbf{V}| - 1)}, \tag{5}$$

i.e., the the ratio of edges actually present in a graph to the number of possible edges [39]. Compared to small-world networks, a random graph typically exhibits a small average minimum path length and small average clustering coefficient [35], depending on the graph's density. In the present study, we created two test cases with $\rho = 0.25$ and $\rho = 0.50$ respectively.

**Performance results for simulated networks.** Running times of the algorithm increased as a function of graph size $|\mathbf{V}|$ and critical path weight $w_{crit}$ (Fig 8). For the dynamic programming part of the algorithm, running times increased in a linear fashion in both network size $|\mathbf{V}|$ (Fig 8A) and critical path weight (Fig 8B). Running times thus correspond to theoretically expected running times. The time needed for backtracking the obtained solutions grew exponentially in $|\mathbf{V}|$ (Fig 8C) and critical path weight (Fig 8D) for random networks and scale-free networks, where running times were least favorable for random networks. For small-world networks on the other hand, running times did not increase dramatically for higher values $|\mathbf{V}|$ and $w_{crit}$.

Running times for backtracking depend on the number of alternative paths to be reconstructed from the solution array. Since the number of paths increases exponentially in $w_{crit}$ (given a sufficient graph size $|\mathbf{V}|$), exponential running times were expected for higher values for both parameters. We therefore defined an a priori limit of 20,000 for the number of alternative paths to be reconstructed. If this limit was reached, the algorithm's execution was aborted. These problem instances were considered intractable (red markers in Fig 8). Intractable test cases were found for random graphs only and occurred earlier for graphs with higher densities of $\rho = 0.50$ (for comparison: the scale-free graphs had a density of approx. 0.15, small-world graphs of approx 0.5). Cases of intractability were found for random graphs of sizes $|\mathbf{V}| \geq 65$ with a density of $\rho = 0.50$ and $|\mathbf{V}| \geq 130$ for graphs with density $\rho = 0.25$ respectively. Furthermore, intractable cases occurred for path weights $w_{crit} \geq 21$ for random graphs with density $\rho = 0.50$ and for path weights $w_{crit} \geq 25$ for random graphs with density $\rho = 0.25$.

Thus, network size as well as network structure influenced the computational demand of a given input to the presented algorithm. Note that intractable cases may well occur in a neuroscience application, where inputs can not be assumed to be bounded in any respect (e.g. in terms of graph density or graph size). Here, the network size may be used to determine whether an input may prove intractable. In the present simulation, network sizes smaller than 25 nodes posed no problem for the algorithm; of course, these limits are subject to moderate changes with increasing computational power.

## Detection of spurious interactions in networks derived from electrophysiological time series

**Ethics statement.** To test the algorithms applicability to biological time series, we used MEG data recorded from 15 healthy human subjects during a face recognition task as described in [40]. All subjects gave written informed consent before the experiment. The study was approved by the local ethics committee (Johann Wolfgang Goethe University, Frankfurt, Germany).

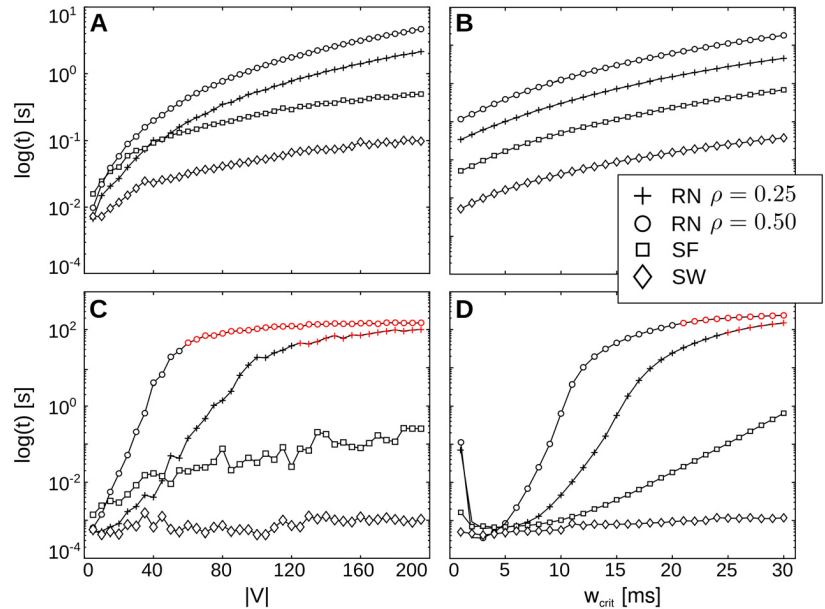

**Fig 8. Results running time.** Running times [log(s)] for dynamic programming (A, B) and backtracking (C, D) by number of vertices |V| and maximum path weight $w_{crit}$. Running times are shown for different graph types (SW: small-world, SF: scale-free, RN: random networks with density $\rho$). Red markers indicate cases of intractability (execution was aborted after a pre-defined limit of reconstructed alternative paths was reached).

doi:10.1371/journal.pone.0140530.g008

**Preparation and MEG data acquisition.** MEG data was obtained from 30 healthy subjects, recruited from the local community. All participants had normal or corrected-to-normal vision and were right-handed (assessed by the Edinburgh Handedness Inventory [41]).

MEG data were recorded using a 275-channel whole-head system (Omega 2005, VSM Med-Tech Ltd., BC, Canada) at a rate of 600 Hz in a synthetic third order axial gradiometer configuration (Data Acquisition Software Version 5.4.0, VSM MedTech Ltd., BC, Canada). Data were filtered with 4th order butterworth filters with 0.5 Hz high-pass and 150 Hz low-pass. Behavioral responses were recorded using a fiber-optic response pad (Lumitouch, Photon Control Inc., Burnaby, BC, Canada). Trials with excessive head movement (more than 5 mm) were excluded from further analysis.

Structural magnetic resonance images were obtained with a 3 T Siemens Allegra, using 3D magnetization-prepared rapid-acquisition gradient echo sequence. Anatomical images were used to create individual head models for MEG source reconstruction.

**Task.** The participants were presented with a randomized sequence of degraded two tone images of human faces (Mooney Faces, [42], see Fig 9C for an example stimulus) and scrambled stimuli, where black and white patches were randomly rearranged to minimize the likelihood of detecting a face. The participants had to indicate the detection of a face by button press. Only trials in which faces were correctly identified entered further analysis.

**Data analysis.** MEG data were analyzed using MathWorks MATLAB (2008b, The Math-Works, Natick, MA) and the open source MATLAB toolboxes FieldTrip (version 2008-12-08; [43]), SPM2 (http://www.fil.ion.ucl.ac.uk/spm), and TRENTOOL [20]. We will briefly describe the applied analysis here, for a more in depth treatment refer to [40].

For data preprocessing, we defined experimental trials from the continuously recorded MEG data. A trial was defined as the epoch from -1000 ms prior to stimulus presentation until 1000 ms after stimulus presentation. Trials contaminated by artifacts (eye blinks, muscle activity, or jump artifacts in the sensors) as well as trials with wrong responses were discarded. Trials were baseline corrected by subtracting the mean amplitude between -500 to -100 ms before stimulus onset.

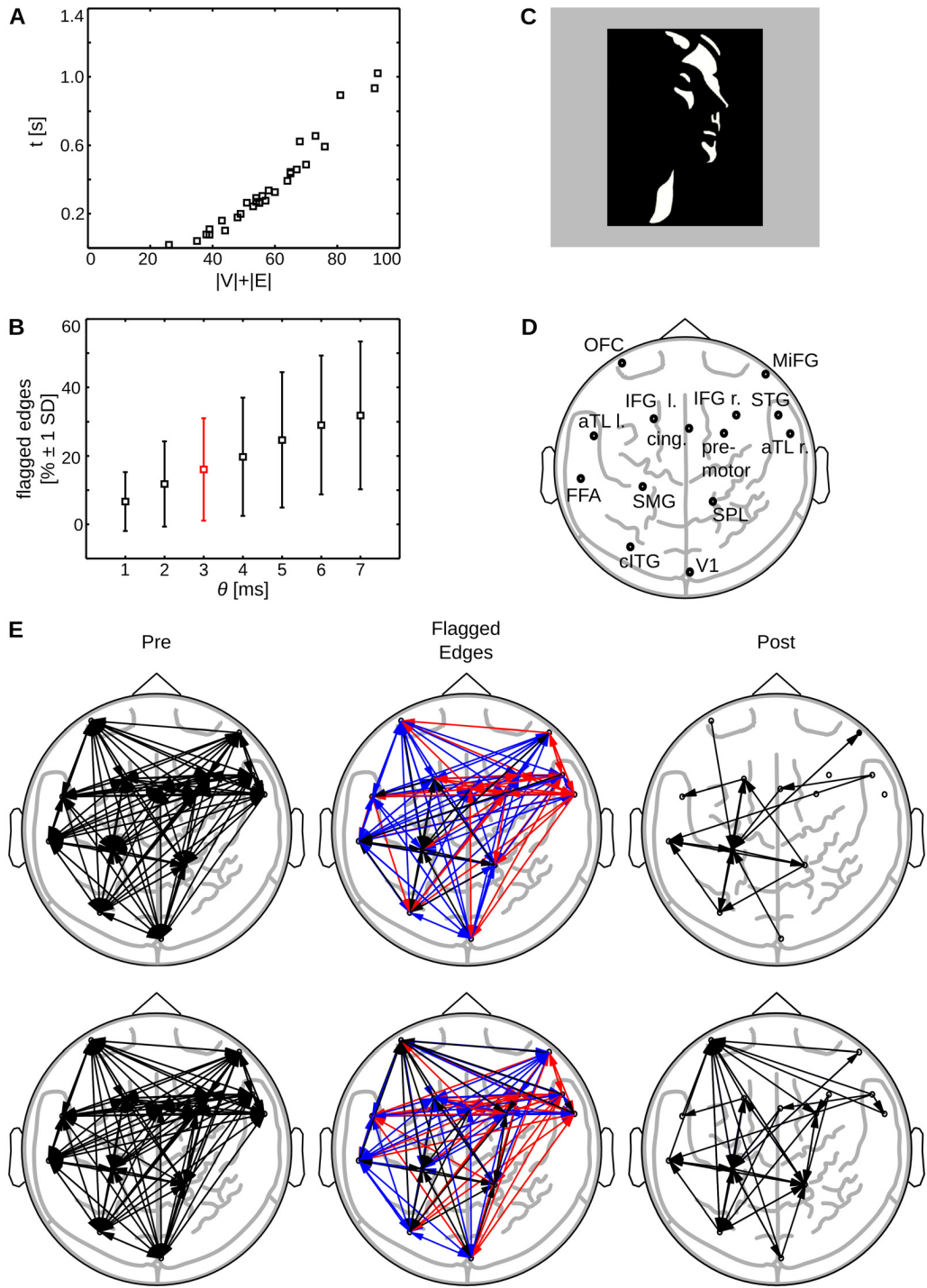

**Fig 9. Results empirical data sets.** (A) Running time of the complete algorithm by number of nodes plus number of edges $|\mathbf{V}| + |\mathbf{E}|$; (B) Mean percentage of tagged, potentially spurious edges by chosen threshold $\theta$ after application of the algorithm, error bars indicate 1 standard deviation (SD); the value for $\theta$ obtained from bootstrapping in two example data sets is marked in red; (C) Mooney Stimulus [42]; (D) Cortical sources after beamforming of MEG data (l.,left; r., right: l. orbitofrontal cortex (OFC); r. middle frontal gyrus (MiFG); l. inferior frontal gyrus (IFG left); r. inferior frontal gyrus (IFG right); l. anterior inferotemporal cortex (aTL left); l. cingulate gyrus (cing); r. premotor cortex (premotor); r. superior temporal gyrus (STG); r. anterior inferotemporal cortex (aTL right); l. fusiform gyrus (FFA); l. angular/supramarginal gyrus (SMG); r. superior parietal lobule/precuneus (SPL); l. caudal ITG/LOC (cITG); r. primary visual cortex (V1)), see also [40]; (E) Example of removal of tagged edges: MEG data of a face detection task in two subjects. First column shows transfer entropy values prior to detection of potentially spurious edges (**Pre**). The second column shows color-coded tagged edges (red: Potential cascade effects; blue: potential common drive effects; $\theta = 3ms$). The third column shows the network of directed interactions after removal of all tagged edges (**Post**).

To investigate differences in source activation in the face and non-face condition, we used a frequency domain beamformer [44] at frequencies of interest identified at the sensor level (80 Hz with a spectral smoothing of 20 Hz). We computed the frequency domain beamformer filters for combined trials ("common filters") consisting of activation (multiple windows, duration, 200 ms; onsets at every 50 ms from 0 to 450 ms) and baseline data (-350 to -150 ms). To compensate for the short duration of the data windows, we used a regularization of $\lambda$ = 5% [45].

To find significant source activation in the face versus non-face condition, we first conducted a within-subject t-test for activation versus baseline effects. Next, the t-values of this test statistic were subjected to a second-level randomization test at the group level to obtain effects of differences between face and no-face conditions; a p-value < 0.01 was considered significant. We identified 14 sources with differential spectral power between both conditions in the frequency band of interest in occipital, parietal, temporal, and frontal cortices (see [Fig 9D](#) and [40] for exact anatomical locations). Namely, our network representing information flow between sources has 14 nodes. We then reconstructed source time courses for TE analysis, this time using a broadband beamformer with a bandwidth of 10 to 150 Hz.

We estimated TE between beamformer source time courses [20, 46] within an analysis window of 500 ms (-50–450 ms) and tested resulting TE values for their statistical significance [20]. We furthermore reconstructed information transfer delays for significant information transfer by scanning over a range of assumed interaction delays from 5 to 17 ms (resolution 2 ms), following the approach in [19] and parameters used in a similar analysis in [46]. We thus obtained a delay weighted, directed network of information transfer during a face recognition task, consisting of 14 nodes and edges with weights in the range from 5 to 17. We then applied the proposed algorithm to the resulting delay-weighted networks of directed interactions. For two example data sets, we used bootstrapping (1000 resampled cases) to obtain an estimate of the standard error of the delay estimation [47] and used this estimated standard error as input parameter $\theta$ for a more detailed example application of the algorithm.

**Performance results for "empirical" networks.** For empirical data running times increased almost linearly in $|\mathbf{V}| + |\mathbf{E}|$ ([Fig 9A](#)). We chose to present running times as function of the sum of the number of nodes and number of edges because a systematic variation of network size was not possible here (rather, network size was determined by previous source reconstruction). Cases of intractability did not occur even though some data sets exhibited high network densities (ranging from 0.07 to 0.43 with a mean of 0.24 and a SD of 0.09).

The percentage of potentially spurious edges increased with higher thresholds $\theta$ up to 32% of potentially spurious edges for $\theta = 7\,ms$ ([Fig 9B](#)). Edges were considered potentially spurious if at least one alternative path existed or a simple CD was present. Note, that the threshold serves to adjust the algorithm's sensitivity and may lead to the erroneous exclusion of edges if chosen too high. Thus, the value for $\theta$ should be chosen such that imprecisions in interaction reconstruction are accounted for, while the false discovery rate is not increased. As a rule of thumb, a user may use prior knowledge about the minimum interaction delay to be expected in the data as an upper bound for $\theta$ or use bootstrapping to obtain an error estimate for reconstructed delays.

In [Fig 9E](#), we show results for two example MEG data sets from the validation test-set before and after analysis with the proposed algorithm. We used bootstrapping to estimate the standard error in the reconstruction of the interaction delays. We found an average error over channels of 2.6 ms for subject one and 2.9 ms for subject two. Accordingly, we set $\theta = 3\,ms$ as this corresponded to the next integer value in ms. The average reconstructed interaction delay was found to be 7.06 ms (SD: 3.57 ms) for subject one and 6.94 ms (SD: 3.32 ms) for subject two. We also calculated the average weight of the shortest path length as the average weight of the shortest path for each node to every other node in the network; the average path length was 6.84 ms for subject one and 6.81 ms for subject two. Note that the graphs are highly connected prior to the application of the algorithm, such that the shortest path between any two nodes consists of just one edge; thus the average path length is close to the average edge weight.

After application of the algorithm, the recovered networks of information transfer consisted of 20 links for subject one and 34 links for subject two; networks showed an overlap of nine edges, which corresponds to an 45% overlap and is 20 times higher than an overlap of 2% expected purely by chance. Thus, the network can be considered highly consistent.

## Discussion

### Algorithmic detection of potentially spurious edges in delay weighted networks

We have presented an algorithm that finds potentially spurious links arising from bivariate analysis of multi-node networks based on interaction timing. The algorithm identifies the most common motifs causing the reconstruction of spurious links, such that identified links can be subjected to further testing, or removed. By removing all potentially spurious edges, the user obtains a sub-network that is guaranteed to contain only non-spurious edges; this improves the validity of the network representation itself as well as the validity of potential subsequent network analysis. The algorithm thus allows the user to find an approximate representation of multivariate interactions in the data, using only bivariate interaction reconstruction and avoiding the computationally heavy problem of an approximately or even fully multivariate approach.

The presented algorithm may be used in neuroscience to post-process any network of reconstructed bivariate interactions, where interactions are directed and weighted by their estimated delays. We demonstrated the application of the algorithm using a reference implementation in MATLAB as part of the open source toolbox TRENTOOL [20].

### Application in neuroscience

Based on findings in [19], we propose to identify spurious links by their characteristic timing signatures in networks of reconstructed bivariate interactions [14]. In particular, we propose that a link is likely to be spurious if an alternative path with identical timing exists (Fig 2).

We assume that a bivariate information transfer between two nodes and a corresponding alternative path constitute a redundant routing of information. Such a redundant routing conflicts with the hypothesis that the brain evolves under the objective of maximizing economy and efficacy [6, 48] while minimizing biological costs [49, 50] (see for example the "save-wire hypothesis" in [51]). We thus argue that any redundant routing of information between two sources of neural activity –with identical timing– would be implausible, given the brain's organizational principles. Therefore, whenever a redundant routing for a bivariate information transfer is found, our rationale implies spuriousness of either the bivariate information transfer *or* the alternative path. Of the two, we consider the bivariate interaction spurious, because (1) spurious bivariate interactions are a likely artifact in bivariate analysis of multi-node networks; and (2) if a bivariate and thereby *direct* means of information transfer between a source and a target existed, the maintenance of a physiologically more costly alternative path of identical information transfer would be unlikely.

Note, that this rationale exclusively applies to neural systems. Also remember, that the algorithm does not tag *all* alternative routings of information, but only those with a certain timing signature; alternative routings with different delays than the bivariate interaction are not considered redundant and are not tagged.

## Treatment of tagged links in neuroscience applications

Our algorithm tags potentially spurious edges to let the user decide if a tagged edge should be ultimately excluded from the network representation. To minimize erroneous exclusions, the user may inform the decision by additional evidence, e.g. previous anatomical or functional findings. If such previous findings do not exist, we recommend the exclusion of tagged edges.

We consider the erroneous rejection of links favorable over erroneous inclusion, i.e., we suggest to rather commit a false negative error if in doubt. We favor false negative errors because in statistical terms, false negatives are considered less severe than false positives (erroneously including a spurious link), as they yield more conservative results. If the user removes all tagged links, the resulting network is guaranteed to contain non-spurious links only, but some links may be missing from the network.

In triangle motifs, the exclusion of all tagged links will definitely lead to false negatives: Here, two links are tagged but the exclusion of both edges is mutually exclusive; more precisely, the exclusion of one of the two tagged edges destroys the motif giving rise to the second, potentially spurious link. This is illustrated in Fig 2C, where the exclusion of link $(v_1, v_t)$ destroys the CE leading to the tagging of edge $(v_s, v_t)$, and on the other hand, the exclusion of $(v_s, v_t)$ destroys the CD leading to the tagging of $(v_1, v_t)$. In triangle motifs, the rejection of both edges thus produces a false negative error; here, prior anatomical or functional evidence are required to decide which of the two tagged edges is non-spurious.

It is further possible to use modeling approaches to test if tagged links are actually present in the network of bivariate interactions. For example dynamic causal modeling (DCM, [52–54]) may be used to test whether a model containing a certain link is favorable given the observed data over a second model missing this link.

## Types of multivariate effects not identified by the algorithm

The correction performed by the presented algorithm is not exhaustive with respect to all types of multivariate interactions potentially occurring in neuroscience data. The interactions not targeted by the algorithm are of two types: (1) more general cases of CD; (2) synergistic effects [17], i.e., combined effects of two or more sources on a third source. In the following we will discuss the conceptual and practical limitations that prevent an exhaustive algorithmic correction for these two types of multivariate interactions.

**Detection of general common drive effects.** The presented algorithm detects *simple* CD in triangle motifs by listing all links with alternative paths of length two. Our algorithm can theoretically be extended to explicitly search for general cases of CD, where two nodes are commonly driven via arbitrarily long cascades of information transfer (Fig 2A).

General CD may be identified by searching for paths of equal summed interaction delays that have a common source and target node. We again assume that equal summed delays hint at redundant and therefore spurious information transfer. It can then be tested if the source node is a common driver for the last and second to last node in one of the two paths by looking at the information transfer delay between these last two nodes. For an example, see Fig 2A, where the bivariate information transfer in the network forms two paths of equal delays connecting nodes $v_0$ and $v_t$: The link $(v_s, v_t)$ is tagged as spurious because its weight corresponds to the difference in the summed path weights $v_0 \xrightarrow{c} v_s$ and $v_0 \xrightarrow{c'} v_t$: $w_{(v_s, vt)} = c - c'$. In this scenario, $v_0$ is a common driver of nodes $v_s$ and $v_t$. This approach also allows to test for higher order CD, i.e., one source driving three or more nodes simultaneously. A similar algorithm was proposed by Marinazzo and colleagues [55] to identify spurious bivariate links using a network of multi-variately reconstructed interactions (see next section).

Even though an extension to general CD is hypothetically possible, its realization is not feasible in practice: The extension requires that for each network node all originating paths of arbitrary length need to be listed. An algorithm fulfilling this task would have an asymptotic running time many times higher than the algorithm presented in this work: For each node in $\mathbf{V}$, $O(|\mathbf{V}| - 1|!)$ paths of arbitrary length and weight exist (in the first step $O(|\mathbf{V}| - 1)$ nodes can be reached from the current starting node, in the second step $O(|\mathbf{V}| - 2)$ nodes can be reached, and so forth). The asymptotic running time of such an algorithm thus amounts to $O(|\mathbf{V}| \cdot |\mathbf{V}| - 1|!)$. Such a factorial running time is commonly not considered feasible in practice and would limit the application to networks of very small size.

**Detection of synergistic effects.**   Synergistic effects describe information that is transferred from a set of sources to a common target, whereby information is combined in a non-trivial fashion [17, 18, 56, 57]. In this case, looking at the set of sources simultaneously provides information about the target that is not obtainable from looking at each source separately. As a toy example, one can think of three nodes implementing a logical XOR operator, where two nodes serve as binary input and the third node serves as output node. Each state of the output node is the exclusive OR of the two previous input states. A bivariate analysis of every pairwise interaction between the three nodes will not detect any significant interaction, because the pairwise mutual information between any two nodes is 0. Analyzing the triplet of nodes simultaneously will however detect an interaction; e.g. the conditional mutual information (TE) will be greater than zero, because it "decodes" the information in one source by conditioning on the second source.

Consequently, synergistic effects between a set of sources and a target node can only be revealed if the whole set is considered simultaneously in some multivariate reconstruction of interactions or by explicitly reconstructing synergistic interactions [18, 58]. Such synergistic effects are not targeted by design of our algorithm as it simply post-processes results from bivariate network analyses. To include synergistic effects, a multivariate interaction analysis would have to replace the estimation of bivariate interactions. Note however, that any fully multivariate method for interaction reconstruction would need to identify the optimal subset of sources that exert some meaningful influence on a given target node. The identification of such an optimal set of sources would require the exhaustive testing of the power set $\mathcal{P}(\mathbf{V})$ of all network sources, due to the non-additivity of information contributions from individual sources (because of redundant and synergistic effects). The power set has size $|\mathcal{P}(\mathbf{V})| = 2^{|\mathbf{V}|}$, i.e., testing all sources brute force has a theoretical running time of $O(2^{|\mathbf{V}|})$. In fact, it has been shown that optimal subset selection in regression is an NP-hard problem [10]. This proof extends to source selection for TE due to the equality of TE with Granger causality for *jointly* Gaussian variables [22]. Thus, the reconstruction of truly multivariate interactions in arbitrarily large networks poses a computationally intractable problem (if $P \neq NP$). In the next subsection we will present approaches that try to approximate fully multivariate methods to circumvent the inherent computational complexity of the problem at hand.

## Comparison to other approximative methods for multivariate network reconstruction

The proposed algorithm provides an approximative method for the inference of networks of multivariate interactions to handle the computational intractability of exact network reconstruction. Methods with a similar purpose have been proposed by various authors. In the following we will review some of these methods and list scenarios that may benefit from the application of our algorithm.

**Multivariate reconstruction of effective networks by Lizier and Rubinov.** Lizier and Rubinov [12] proposed a greedy algorithm which for each network node $Y$ (the target) infers a set of influential source variables $\mathbf{V_Y}$. A source is considered influential if it adds significantly to the information transfer from $\mathbf{V_Y}$ to $Y$. The set $\mathbf{V_Y}$ is thus built by iteratively adding sources, which have significant information transfer into $Y$, conditional on the previously included sources. Finally, information transfer is re-evaluated conditional on the complete set of included sources:

$$TE(X \rightarrow Y | \mathbf{V_Y}) = I(X; Y | \mathbf{V_Y} \setminus X),$$

i.e., the mutual information between each source $X \in \mathbf{V_Y}$ and target $Y$ while conditioning on all remaining relevant sources in $\mathbf{V_Y}$, except $X$. If a source fails to provide statistically significant information about the present of $Y$, it is removed from $\mathbf{V_Y}$. After this "pruning step", the set $\mathbf{V_Y}$ consists of all relevant sources that contribute information about the target. The approach is robust against the detection of spurious interactions due to CD and CE, because for each interaction reconstruction it conditions on all relevant sources in the network.

Note that the greedy strategy used by Lizier and Rubinov is approximative insofar as it does not guarantee a maximal informative set $\mathbf{V_Y}$ over all sets $\mathcal{P}(\mathbf{V})$. For example, purely synergistic effects between two or more sources may be missed. The authors propose to extend their greedy method by also testing tuples and higher order combinations of sources, but they note that this requires considerable computational resources, which may not be worth the gain in information. The testing of tuples may however be feasible for small networks.

**Partial conditioning of information transfer by Marinazzo and colleagues.** Marinazzo and colleagues [55] proposed a greedy algorithm resembling the approach by Lizier and Rubinov. Again, the algorithm tries to account for other relevant network sources when evaluating the information transfer from a source $X$ to a target $Y$ in a multivariate system. To identify relevant sources, Marinazzo et al. propose to iteratively construct a "partial conditioning set" $\mathbf{Z}$ from all sources $\mathbf{V} \setminus \{X, Y\}$. In each iterative step $k$ the algorithm includes the source $Z_k$ that maximizes the mutual information between $\mathbf{Z_k}$ and the source $X$, i.e., $Z_k = \max_Z (I(X; \mathbf{Z_k}))$, where $\mathbf{Z_k} = \mathbf{Z_{k-1}} \cup Z$.

The partial conditioning approach may miss interactions if sources share a lot of redundant information about a target: If an existing source-target relationship is evaluated while conditioning on sources providing redundant information about the target, the source-target relationship under investigation is not detected. Therefore, Stramaglia and colleagues [59] extended partial conditioning with a graph algorithm that identifies these missed interactions. The authors proposed to reconstruct interactions multivariately (e.g. with partial conditioning) and bivariately. The multivariate network is then used to algorithmically separate bivariate links in two sets: (1) links explained by an indirect path of information transfer in the multivariate network (CE), and (2) links not explained by an indirect path. Bivariate links in the second group, which are missing from the multivariate network, are assumed to reflect non-spurious information transfer that was missed by the multivariate approach. These bivariate links are then merged with the multivariate network.

Note that the rationale underlying the algorithm proposed in [59] resembles the rationale presented in this paper because spurious bivariate interactions are identified by alternative paths; however, the aim of both approaches differs: The algorithm presented by Stramaglia and colleagues improves multivariate interaction reconstruction, while the algorithm presented in this paper tries to approximate a multivariate approach from bivariate interaction reconstruction alone. The approach proposed by Stramaglia thus still requires the potentially intractable reconstruction of multivariate interactions from data.

**Non-uniform multivariate embedding by Faes and colleagues.** Faes and colleagues [60] proposed a non-uniform embedding technique to estimate the information transfer from one source variable to a target in the presence of further potentially relevant sources of information transfer. The authors propose to iteratively build a non-uniform embedding vector from a set of candidate time points from the past of all sources in a network up to a certain predefined limit. Points are included in the vector if they add significant information about the next state of the target. Information transfer between a source and a target may then be estimated while conditioning on this non-uniform embedding vector.

The iterative construction of the embedding vector follows a greedy strategy that is similar to the strategies discussed above [12, 55]. Accordingly, the returned embedding vector is not guaranteed to yield the set of maximally informative source time points with respect to the target, as it will miss purely synergistic contributions of two or more points to the target. As said above, an exhaustive testing of all possible subsets of source time points poses an intractable computational problem. This is explained next.

**Exhaustive brute force analysis.** A brute force analysis of interactions between all possible subsets ($m$-tuples) in a set of nodes would yield an exact solution to the problem of inferring the network of multivariate interactions from data. For the example of $m = 3$, one would enumerate all 3-tuples or triplets in the set of nodes and for each triplet evaluate the six possible interaction motifs—three potential targets and for each target two possible combinations of source and conditioning node. Note that here the mandatory conditioning takes care of potential synergies. For the general case of $m$-tuples, this would generalize to $\binom{|V|}{m}$ possible subsets of size $m$, where for each tuple $m \cdot (m - 1)$ possible interaction motifs exist. As for approximative approaches, such an analysis is feasible for small numbers of $m$ only.

## Application scenarios for the proposed algorithm

The most basic application scenario for the proposed algorithm is as post-processing step after bivariate reconstruction of directed, delay-weighted interactions from a set of neural sources. Here, the algorithm helps to prune potentially spurious edges to obtain an approximative, statistically conservative network representation of the physical interactions. In this scenario, our algorithmic correction is favorable over multivariate interaction reconstruction whenever available data is limited or high-dimensional, such that data points are not sufficient to estimate highly multivariate interactions. Here, our algorithm is more data-efficient because it relies on bivariate interaction reconstruction only. Such data-efficiency is especially relevant for information theoretic measures, where quantities are often estimated using kernel estimators or neighbor methods (as for example proposed in [61]); applying these kernel estimators to the estimation of highly multivariate data may lead to very high dimensional search spaces, which suffer from the curse of dimensionality, hindering a reliable estimation of the quantities of interest.

The application of our algorithm may prove beneficial prior to calculating graph theoretical measures from networks of reconstructed interactions. We argue that these measures are more reliable when applied to a statistically conservative network representation.

The algorithm may further be used in conjunction with modeling approaches such as DCM, where it serves to limit the model space to be tested. DCM may also help to identify the most plausible network representation from models, after in- and excluding individual tagged edges respectively.

The presented algorithm may further serve as a preprocessing step for trivariate estimation of information transfer: By testing only the triangle motifs identified by the algorithm, the number of necessary information transfer estimations reduces drastically compared to the brute-force approach discussed in the last subsection. Necessary estimations include bivariate interaction reconstruction ($|\mathbf{V}| \cdot (|\mathbf{V}| - 1)$ calculations for $|\mathbf{V}|$ nodes) and subsequent multivariate interaction reconstruction for identified triangle motifs. The actual number of multivariate reconstructions depends on the number of motifs: The approach is asymptotically faster if 90% or less of the possible $\binom{|\mathbf{V}|}{3}$ triangle motifs are actually present in the data (for network sizes $|\mathbf{V}| > 12$). Trivariate estimation of TE has been implemented in TRENTOOL [20], which also includes the reference implementation of the proposed algorithm. Thus, both analyses may be used in conjunction to estimate multivariate TE of order three. An extension to higher orders is theoretically possible although not implemented as it is not deemed feasible for practical purposes.

Finally, the algorithm is especially suitable for the application in simulated networks where all information transfers have a delay of unity, such as elementary cellular automata, and spurious interactions are therefore easily found.

## Author Contributions

Conceived and designed the experiments: PW UM MW. Performed the experiments: PW. Analyzed the data: PW MW. Contributed reagents/materials/analysis tools: PW MW UM. Wrote the paper: PW MW. Designed the algorithm used in analysis: MW PW UM. Implemented the algorithm used in analysis: PW.